\documentclass[a4paper,fleqn,usenatbib]{mnras}

\usepackage{mathptmx}
\usepackage[T1]{fontenc}
\usepackage{ae,aecompl}
\usepackage{multirow}
\usepackage{graphicx}	
\usepackage{amsmath}	
\usepackage{amssymb}	
\usepackage{color} 
\usepackage{setspace}
\usepackage{latexsym,pdflscape,lscape,epsf,ulem}
\usepackage{appendix}
\usepackage{CJK}
\usepackage{mathrsfs,amsfonts} 

\title[Orbital decay of a heartbeat star system]{Searching for orbital decay in a heartbeat star system KIC 3766353} 

\author[J. W. Ou et al.]{Jian-Wen Ou,$^{1,2,3}$ \thanks{E-mail: oujw3@mail.sysu.edu.cn}
Cong Yu,$^{1,2}$ \thanks{E-mail: yucong@mail.sysu.edu.cn}
Chen Jiang,$^4$
Ming Yang,$^{3,5}$
and Hubiao Niu$^{6,7}$
\\
$^{1}$ School of Physics and Astronomy, Sun Yat-sen University, Zhuhai 519082, China;\\
$^{2}$ CSST Science Center for the Guangdong-Hongkong-Macau Greater Bay Area, Zhuhai 519082, China;\\
$^{3}$ Key Laboratory of Modern Astronomy and Astrophysics in Ministry of Education, Nanjing University, Nanjing 210023, China;\\
$^4$ Max-Planck-Institut f\"ur Sonnensystemforschung, Justus-von-Liebig-Weg 3, G\"ottingen 37077, Germany;\\
$^5$ School of Astronomy and Space Sciences, Nanjing University, Nanjing 210023, China;\\
$^6$ Department of Astronomy, Beijing Normal University, Beijing 100875, China;\\
$^7$ Xinjiang Astronomical Observatory, Chinese Academy of Sciences, Urumqi 830011, China.\\
}

\date{Accepted 2021 September 24. Received 2021 August 30; in original form 2021 May 8.}

\pubyear{2019}

\begin{document}
\label{firstpage}
\pagerange{\pageref{firstpage}--\pageref{lastpage}}
\maketitle

\begin{abstract}

Theory suggests that the orbits of a large fraction of binary systems, including planet-star binary systems, shrink by few orders of magnitude after formation. But so far, only one hot Jupiter with tidally-driven orbital decay has been found by transit timing variations. We propose to search for orbital decay companions  in heartbeat star systems because the orbital angular momentum is effectively transferred to the host star causing tidal dissipation. KIC 3766353 is one of the heartbeat stars with tidally excited oscillations. We acquired the primary and the secondary eclipse time variations from the \textit{Kepler} photometric light curves. Timing analysis shows that KIC 3766353 is a hierarchical triple system with a hidden third body and a red dwarf (mass $0.35\ M_{\odot}$, radius $0.34\ R_{\odot}$) in its inner orbit. The minimum mass of the third body is $\sim 0.26 \ M_{\odot}$, and the distance from the inner orbital is $\sim 111.4 \ R_{\odot}$. The period decay rate of the red dwarf is approximately 358 ms yr$^{-1}$. The combined effects of the light-travel time and the orbital decay lead to the observed timing variations. Future monitoring with a long time base-line observations is required to  delve into the contributions of these two effects.

\end{abstract}


\begin{CJK}{UTF8}{gbsn}

\section{Introduction}\label{sec1}

The change of orbital period is an elegant indicator of the orbital evolution in binary systems. From an observational point of view, the eclipse time variations (ETVs) via observed-minus-calculated (O$-$C) diagrams has been a traditional approach to investigating the period changes in eclipsing binary systems. There are several different physical processes acting on the changes of orbital periods leading to ETVs. One of the well-known effects is the apsidal motion \citep{Ste1939, Cla1993} which is a gradual shift in the position of periastron. On secular timescales, a stellar companion with eccentric orbit induces periastron precession due to the non-spherical symmetry of the gravitational field from tidal or rotational bulges, as well as the space distortions of the general relativistic correction. Additional apparent orbital period variations may arise from the light-travel time effect (LTTE) which is caused by the projection distance changes of a binary in a hierarchical multiple-star system. The LTTE effect is often used to search for or to verify the presence of a tertiary companion \citep{mon2010, zas2016}. There are also the long-term physical effects that are expected to connect to a real orbital period variation, such as tidal dissipation \citep{Bar2020, LiG2020}, stellar merger \citep{pej2017,pas2019}, or even gravitational radiation \citep{fau1971}. These phenomena cause orbital decays. Apart from the extremely cases, the ETVs manifest as a constant rate change of the orbital period that produces a quadratic O$-$C diagram \citep{Nan2011, Nan2015}.

The tidal interactions between the components in close binaries, including planet-star binary system, are important in the details of the orbital evolution mechanism \citep{maz2008,alv2021, mal2021}. At least two mechanisms, such as the tidal friction and the Kozai cycles \citep{koz1962}, cause the orbits of binary stars and exoplanets to shrink by 1-2 orders of magnitude after formation \citep{fab2007}. The tidal dissipation is expected to lead to the orbital circularization \citep{van2016, nine2020}, the spin-orbit circularization \citep{lur2017}, and the spin-orbit misalignment \citep{ogi2014, lin2017}. Recently, the exciting indications of tidal dissipation that results in orbital decay where first confirmed in a short-period hot Jupiter planetary system WASP-12 based on transit timing variations \citep{Yee2020, Tur2021}. The measured decay rate of the orbital period provides an opportunity to gain an insight into the tidal mechanisms in stars \citep{Mac2016,Pat2017,Wil2017,pet2018}.

Orbits of short period binary systems with sufficiently massive secondaries ($\geqslant 10^{-2} M_{\odot}$) should be decaying rapidly. This is true especially for the primary stars with surface convective zones that effectively convert the orbital angular momentum through tidal interactions induced by the massive companion \citep{sun2018, Bar2020}. There are some evidences of tidal evolution in eccentric binaries containing low stellar mass component \citep{tria2017} and in evolved stars \citep{beck2018,pric2018}. From an energy point of view, the orbital decay requires the transfer of angular momentum from the orbits to stars. On one hand, the frictional processes within the stars damp the orbital motions, moving the companion toward the host star. On the other hand, the converted energy of angular momentum causes the host star to spin faster, to heat the internal of the components, and even to excite the stellar resonant oscillations, i.e tidally excited oscillations \citep[TEOs,][]{Ful2017}. Heartbeat stars are the prototype of tidally excited oscillations, which are  undergoing periodic tidal forces in close binary systems with eccentric orbits \citep{Tho2012}. Hence, such stars appear to merit a high priority for searching orbital decays that are potentially detectable. Therefore, in this paper, we propose an analysis to explore the orbital decay in one of the heartbeat star systems KIC3766353.

KIC 3766353 is classified in Kepler Eclipsing Binary Catalog\footnote{http://keplerebs.villanova.edu} \citep{prsa2011}. This particular system was initially identified as an exoplanet candidate\footnote{https://exoplanetarchive.ipac.caltech.edu/overview/KOI-6359}, but later was determined to be an eccentric binary with heartbeat signatures \citep{kirk2016}. In previous comprehensive survey projects, KIC 3766353 were also analysed by \cite{Con2014} and \cite{Bor2016} to search for ETVs. Both analyses implied that a potential third body companion exist in this system. Considering the particularity of this individual target, i.e. the presence of tidally excited oscillations, KIC 3766353 is likely to have orbital decay due to the orbital angular momentum transfer.

The structure of this paper is organised as follows. The photometric data of KIC 3766353 is described in Section \ref{sec2}. In Section \ref{sec3}, we solve the orbital and stellar parameters by using the PHOEBE package for modeling the peculiar light curve. Section \ref{sec4} presents the process of timing analysis. The measurement of mid-times of primary and secondary eclipses is described in Section \ref{sec4.1}, and the correlation of timing variations is presented in Section \ref{sec4.2}. In Section \ref{sec5}, we discuss the contributions of orbital decay (Section \ref{sec5.1}), LTTE effect (Section \ref{sec5.2}), and dynamical perturbation (Section \ref{sec5.3}) to the timing variations of KIC 3766353, respectively. The implications of ETVs are drew in Section \ref{sec_add6}. Finally, we summarize our findings in Section \ref{sec6}.

\section{Photometric data of KIC 3766353}\label{sec2}

The data used for the analysis in this paper are the \textit{Kepler} quarter (Q) 1 to 17  long-cadence (29.4 min) data. In total we have 13 quarters of \textit{Kepler} data, because the target was not observed in Q0, Q6, Q10 and Q14. We also note that short-cadence (58.89 s) data not available for this star. 

\begin{figure*}
\centering
  \includegraphics[width=0.95\textwidth]{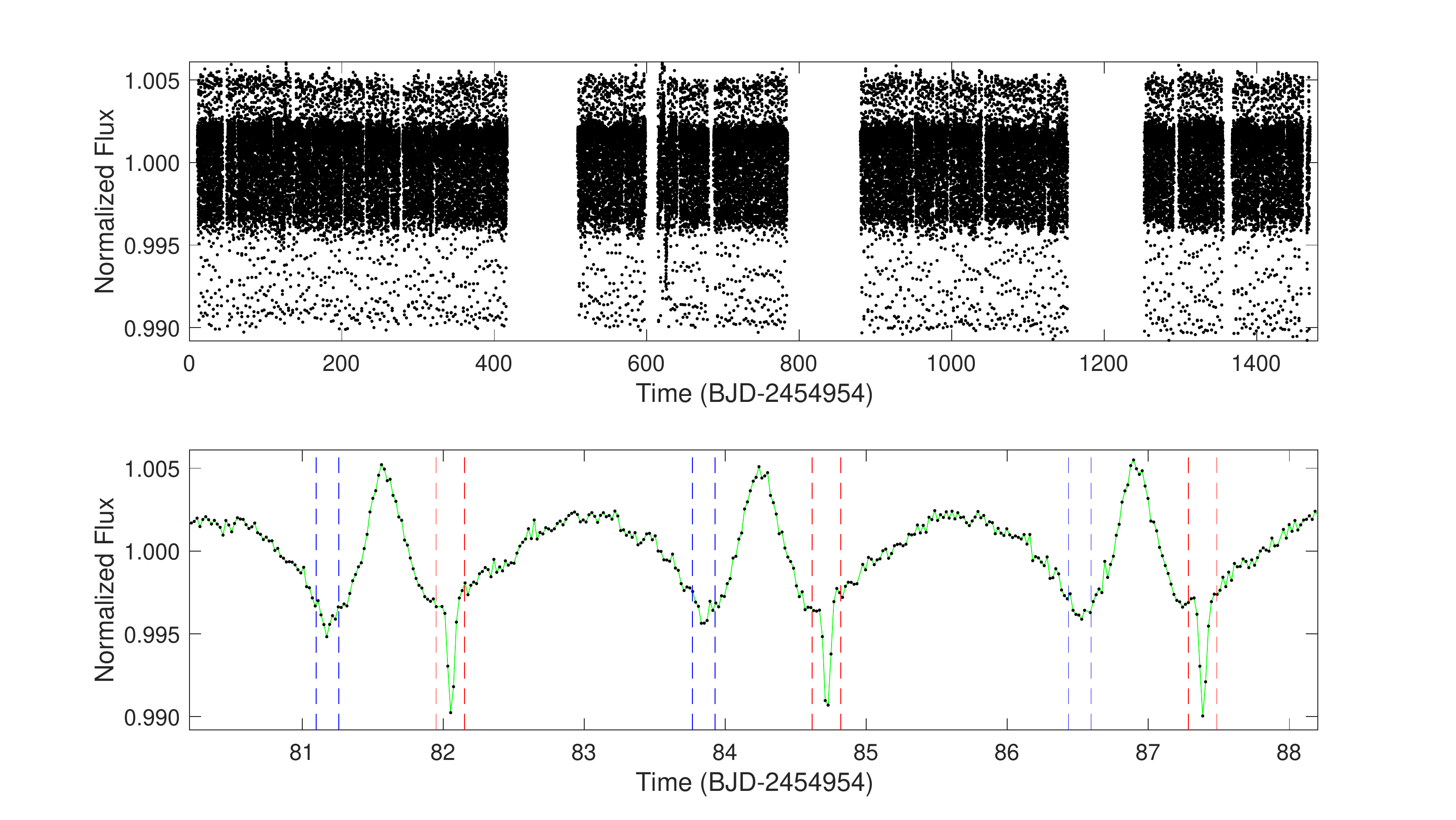}
  \caption{Top panel: the normalized flux of KIC 3766353 for Q1-Q17. The three large gaps are due to data missing for Q6, Q10 and Q14. Bottom panel: a zoomed-in image of three successive eclipses. The vertical dashed lines are near the ingress or egress, red and blue represent the primary and secondary eclipses, respectively. The limb-darkening of ingress and egress are strongly affected because of the stellar oscillations.}
  \label{fig1}
\end{figure*}

In order to remove the long-term systematic trend, we detrended and normalized the original light curve by fitting a low-order ($\leqslant$6th) polynomial to individual segments of data separated by \textit{Kepler} observational gaps. Only the out-of-eclipse light curve was elected to fit the polynomials, and the specific order was chosen as the one that minimizes the standard deviation. The outliers of all data points were cliped by a 5$\sigma$ criterion. In Figure \ref{fig1} the top panel shows the normalized flux from May 2009 to May 2013, and the bottom panel is a zoomed-in image for the data covering three successive eclipses that show remarkable heartbeat-like features. 

It is worth noting that, heartbeat stars oscillate throughout their orbits due to the periodic tidal forces \citep{Ful2017}. Therefore, the ingress and the egress of the primary and the secondary eclipses are contaminated by the stellar oscillations. An illustrative example is the three successive eclipses shown in the bottom panel of Figure \ref{fig1}. Therefore, the precision of ETVs is reduced if the part of limb-darkening is considered, see Section \ref{sec4.1} for details. In short, the special nature of eclipse and heartbeat signals makes KIC 3766353 extremely complex and interesting.

\begin{figure*}
\flushleft
 \includegraphics[width=1.080\textwidth]{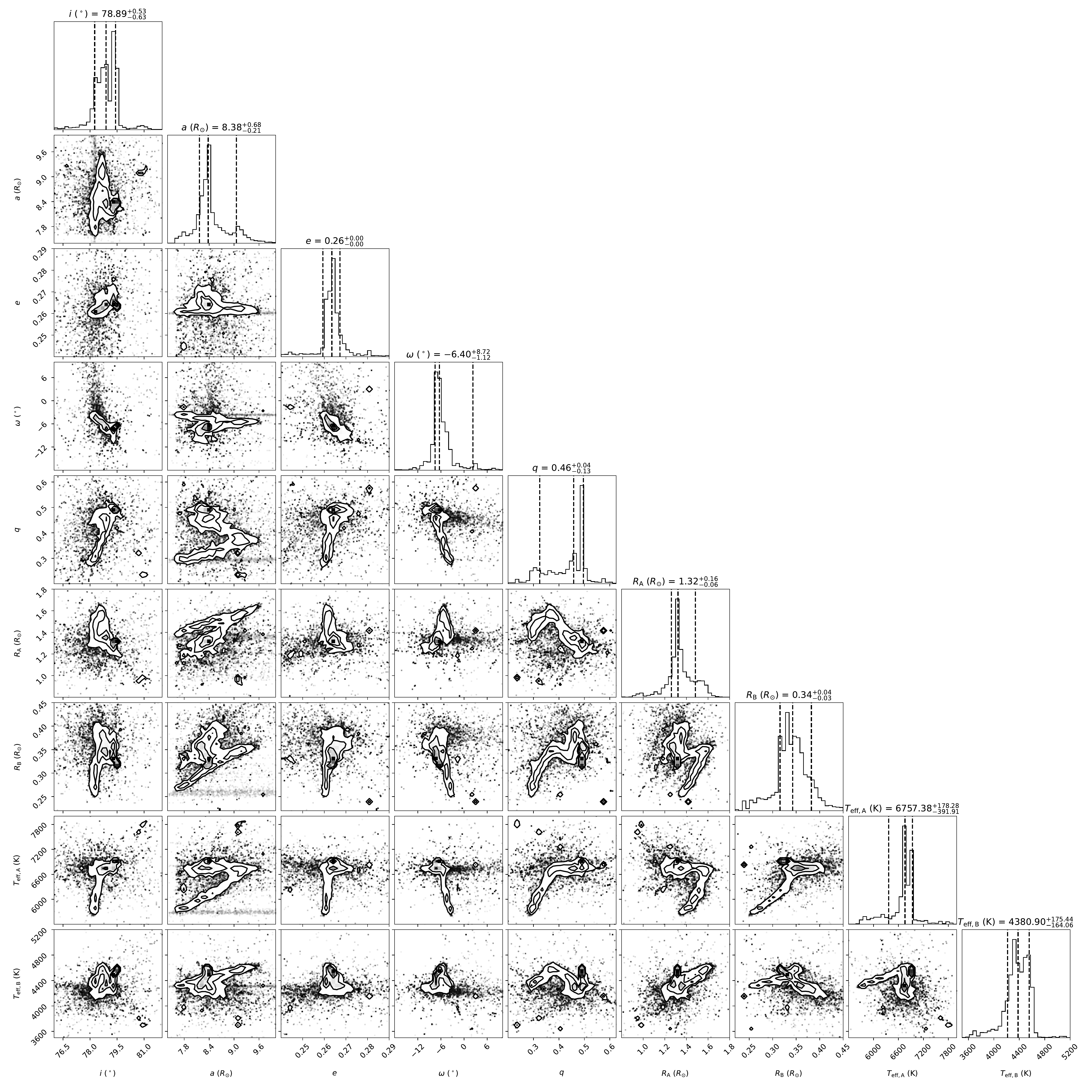}
  \caption{Parameters of the best-fitting model of the binary light curve. Corner distributions depict the two-dimensional projections of the posterior probability of the 9 adjustable parameters. In the one-dimensional distributions, the 16\%, 50\%, and 84\% percentile are marked by the vertical dashed lines. The uncertainties presented above the 1-D distributions correspond to 68\% percentile. }
\label{fig3_mcmc}
\end{figure*}

\begin{figure*}
\centering
  \includegraphics[width=0.95\textwidth]{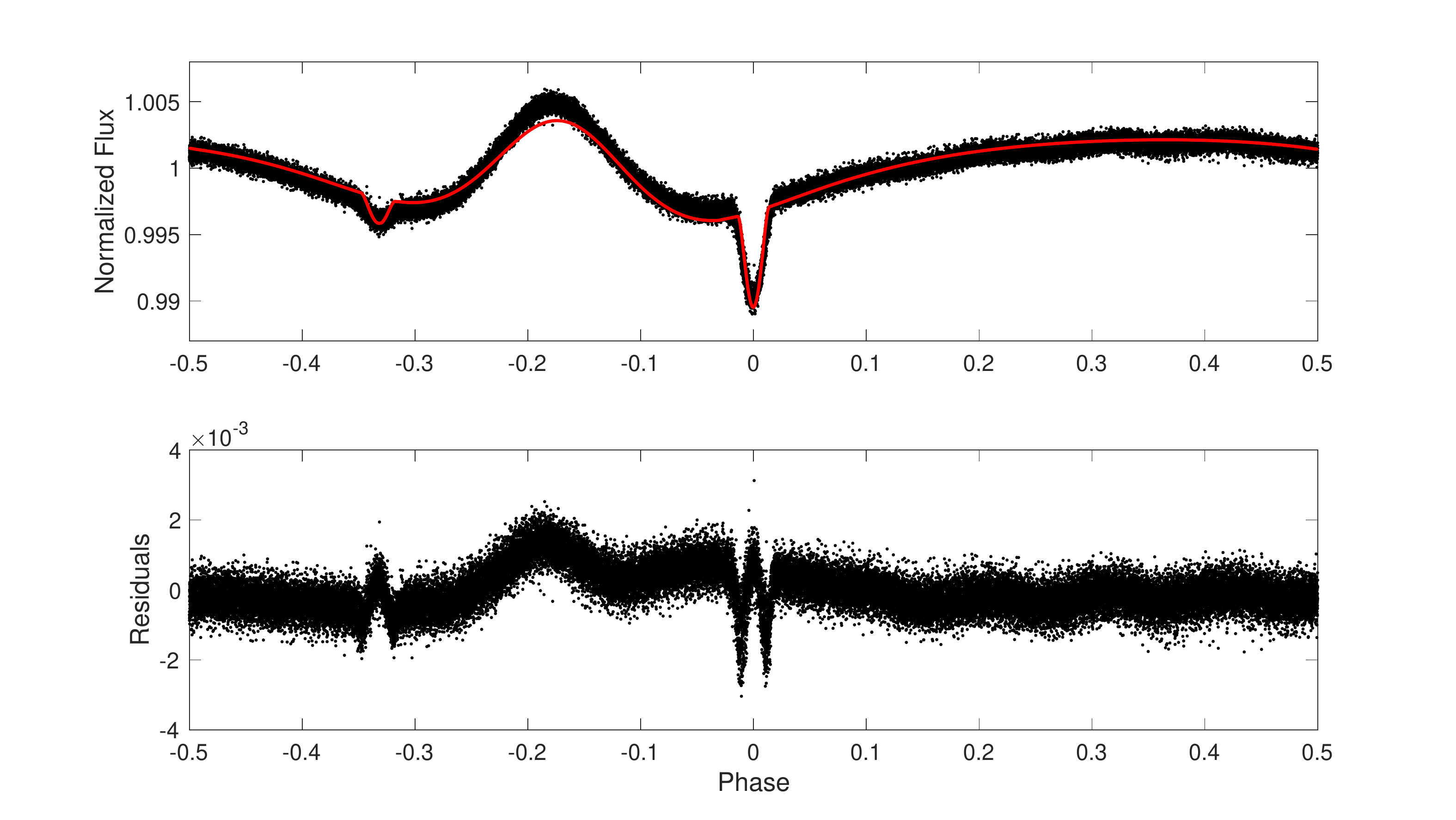}
  \caption{Top panel: phase curve of KIC 3766353 with the PHOEBE's best-fitting model (red solid line). The heartbeat signature appears at the phase of about $-0.2$. Bottom panel: the fitting residuals do not display well in the eclpse parts, which imply that ETVs are present in KIC 3766353.}
  \label{fig2_lcmodel}
\end{figure*}

\section{Orbital and companion parameters}\label{sec3}
We use the python package PHOEBE \citep[version 2.3, ][]{con2020} that adopts the Roche model for surface distortion induced by tidal forces (as well as ellipsoidal variations) to parameterize the orbital and stellar parameters of KIC 3766353. 

During the fitting process, the atmosphere of each component is treated as a blackbody, and the logarithmic limb-darkening law coefficients from \cite{cla2011} are automatically updated. The gravity-darkening coefficients and the bolometric albedos are adopted as default values, 0.32 and 0.6, respectively. The orbital period obtained from the Lomb-Scargle periodogram is fixed.

The adjustable 9 parameters are: the orbital inclination ($i$), the semi-major axis ($a$), the orbital eccentricity ($e$), the argument of periastron ($\omega$), the mass ratio ($q$), the stellar radii ($R_{\rm A}$ and $R_{\rm B}$) and the effective temperatures ($T_{\rm eff,A}$ and $T_{\rm eff,B}$) of both stars with subscript A being the primary and B being the secondary. The fitted parameters and the corresponding uncertainties  are estimated by using the Markov Chain Monte Carlo (MCMC) method implemented in PHOEBE via the  \textsc{emcee} sampler \citep{for2013}. The uncertainties are measured as the 68\% distribution of the parameter scatter, which correspond to one standard deviation from the mean. The geometrical configurations that reproduce the observations best are illustrated in Figure \ref{fig3_mcmc}, in a corner plot manner. The orbital and the stellar parameters of this best-fitting model are also listed in Table \ref{tab1}. The resulted light curve is shown in Figure \ref{fig2_lcmodel}.

Judging from the masses (A: 0.76 $M_\odot$, B: 0.35 $M_\odot$) and the radii (A: 1.32 $R_\odot$, B: 0.34 $R_\odot$) of the two components, we argue that the primary star is possibly a subgiant that evolves just off the main sequence, and the secondary is a red dwarf.

\begin{table}
\footnotesize
\renewcommand\arraystretch{1.5}
  \caption{Orbital and stellar parameters of KIC 3766353.}
  \label{tab1}
  \begin{center}
  \begin{tabular}{lcc}
  \hline
 Orbital parameters & \multicolumn{2}{c}{System}  \\
 \hline
 Orbital period (d): $P_{\rm orb}$ & \multicolumn{2}{c}{$2.666985\pm0.000004$} \\
  Time of primary minimum (d): & \multicolumn{2}{c}{$2455076.680574\pm0.000002$} \\
 Semi-major axis ($R_\odot$): $a$ & \multicolumn{2}{c}{$8.38_{-0.21}^{+0.68}$} \\
 Orbital eccentricity: $e$ & \multicolumn{2}{c}{$0.264 \pm {0.004}$} \\
Argument of periastron ($^\circ$): $\omega$ & \multicolumn{2}{c}{$-6.40_{-1.12}^{+8.72}$} \\
Orbital inclination ($^\circ$): $i$ & \multicolumn{2}{c}{$78.89_{-0.63}^{+0.53}$} \\
  \hline
Stellar parameters & Primary & Secondary\\
\hline
Mass ($M_{\odot}$) & $0.76^{+0.21}_{-0.13}$ & $0.35_{-0.16}^{+0.13}$\\
Mass ratio ($q$) & \multicolumn{2}{c}{$0.46_{-0.13}^{+0.04}$} \\
Radius ($R_{\odot}$) & $1.32^{+0.16}_{-0.06}$ & $0.34_{-0.03}^{+0.04}$\\
$T_{\rm eff}$ (K) & $6757^{+178}_{-392}$ & $4381_{-164}^{+175}$\\
log $g$ (cgs) & $4.08^{+0.22}_{-0.11}$ & $4.92_{-0.27}^{+0.26}$\\
Bolometric albedo (default) & 0.6  & 0.6\\
Gravity brightening (default) & 0.32 & 0.32\\
Limb darkening coeff. (logarithmic) &  auto & auto \\
  \hline
\end{tabular}
\end{center}
\end{table}

\section{Timing Analysis}\label{sec4}
\subsection{Measuring mid-eclipse times of primary and secondary}\label{sec4.1}

The measurement of mid-eclipse time is strongly affected by the symmetry of the photometric data. The asymmetric structure of each eclipse can be corrected by a non-linear model. To do this, we apply a robust ``loess" smooth method to the light curve that is phase folded by its orbital period. Only the out-of-eclipse parts are consider, and the window size of the smooth method is chosen to be 0.186 d that is determined according to  the minimum of the standard deviation between the observed and  smoothed data. Finally, the horizontal light curve with symmetric eclipse is obtained by calculating the relative fluxes between the observed and smoothed light curve. As shown in Figure \ref{fig_etv}, the unprocessed asymmetric eclipse data are shown in the top panels, and the horizontal light curves with symmetric eclipse are shown in the bottom panels. The left is the primary, and the right is the secondary.

The limb-darkening of ingress and egress suffers from stellar oscillations as well, see the bottom panel of Figure \ref{fig1}. The precision of mid-eclipse times is significantly reduced if a theoretical binary model containing ingress and egress parts is used. Therefore, we decide to measure the mid-eclipse time and its uncertainty by a polynomial template \citep{Rap2013, Yang2015}. Only the eclipse part of the symmetric light curve is considered. The template does not represent a physical model, but rather mathematically searches a local minimum flux on the symmetric light curve, of which the time  represents accurate mid-eclipse. The polynomial template is:

\begin{figure}
\centering
  \includegraphics[width=0.5\textwidth]{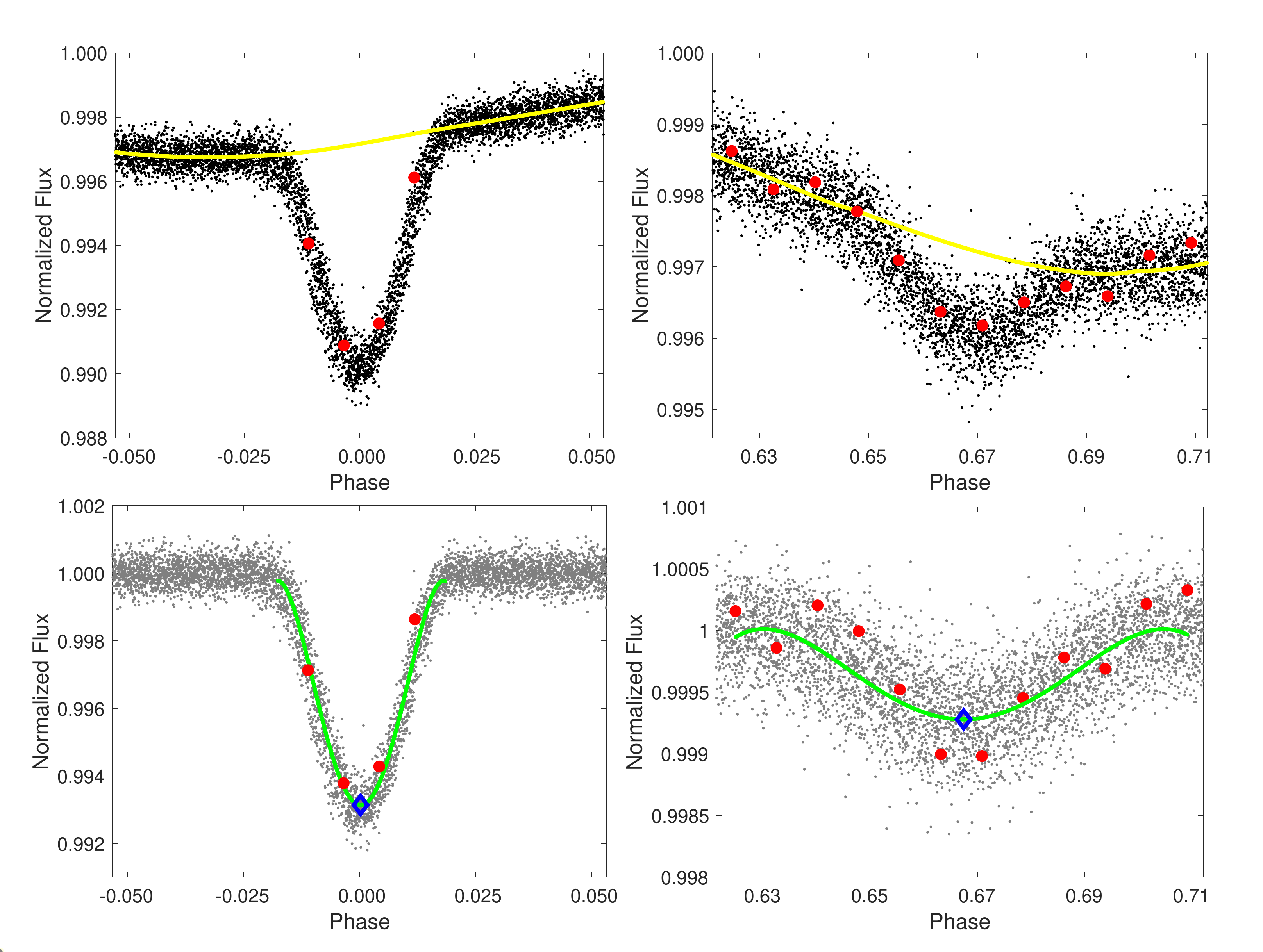}
 \caption{The top panels are the unprocessed data with asymmetric eclipse part that affects the measurement of the mid-eclipse time, while the horizontal light curve with symmetric are showed in the bottom panels. The yellow line is a smooth line for the out-of-eclipse part. The big red dots represent the observed data of an arbitrary individual eclipse that are used for fitting the polynomial template (green line). The blue diamond is the best-fitting time $t_0$ of the mid-eclipse for primary (left panel) and secondary (right panel), respectively.}
  \label{fig_etv}
\end{figure}

\begin{equation}
F=\alpha (t-t_0)^2 + \beta (t-t_0)^4 + F_0 \ , \label{equ1}
\end{equation}
where $F$ is the flux, $F_0$ is the minimum flux, $t$ is time, and $t_0$ is the mid-time. First, we obtain the coefficients $\alpha$, $\beta$ and $F_0$ of the theoretical template by fitting the symmetric phase  curve of all observation data. Then, the whole light curve is divided into many individual segments for each epoch, i.e. the number of orbits. For a single event, the mid-time $t_0$ is set to be a free parameter, but all the other parameters are fixed to be the same as the theoretical template. Finally, the fitted polynomial parameter $t_0$ documents the observed time of mid-eclipse in each individual segment. The segments that do not cover a whole eclipse part are excluded from the fitting, because such segments introduce large errors.

\begin{figure}
\centering
  \includegraphics[width=0.5\textwidth]{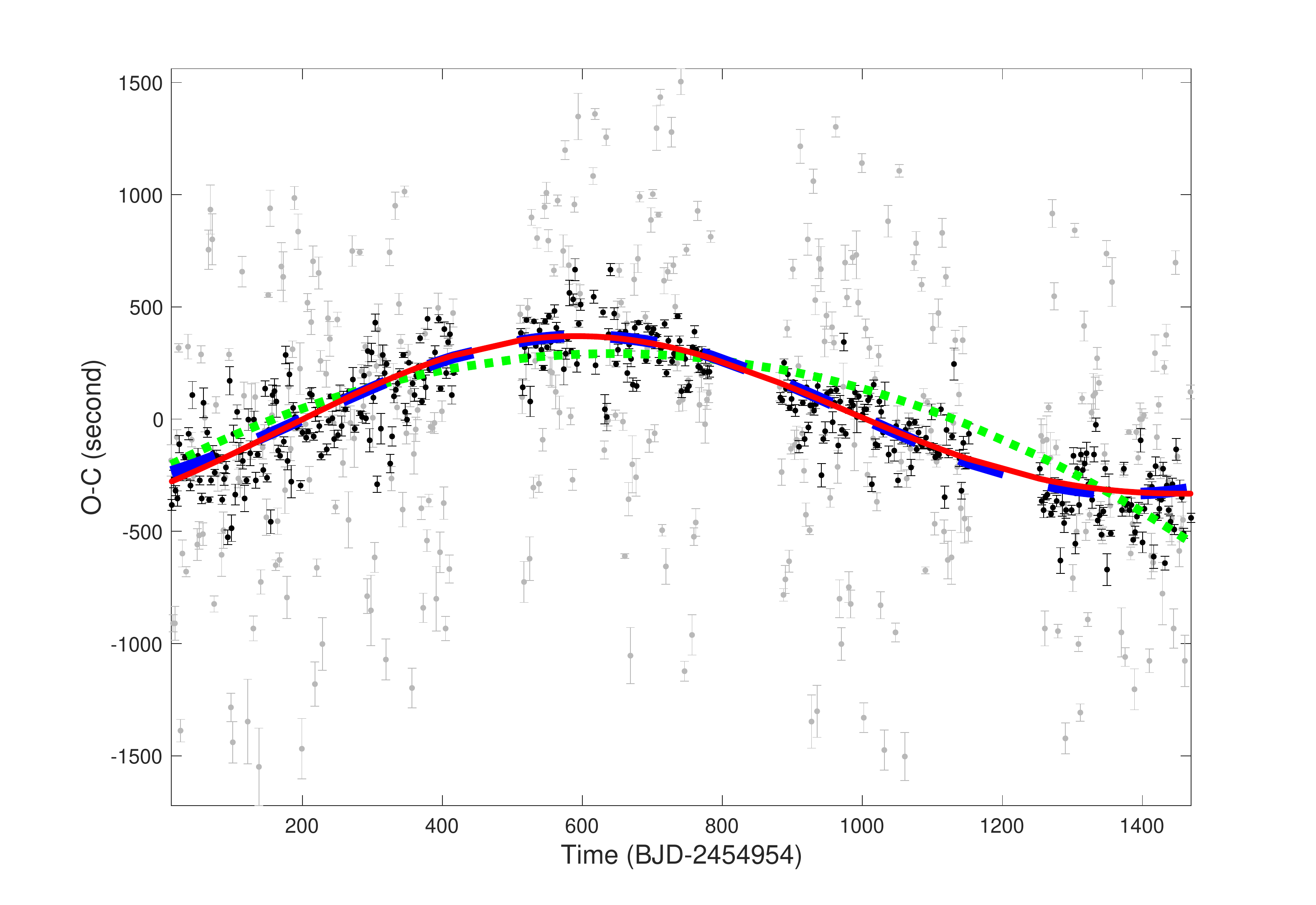}
 \caption{Eclipse timing variations of KIC 3766353. Each data point is the difference of mid-time between observed data and calculated curve, i.e. O-C. The black and gray dots with error bars represent the primary and secondary ETVs, respectively. The best-fitting models of the pure orbital decay (Equation \ref{equ_dec}), the pure LTTE (Equation \ref{equ_ltte}) and the combined model ($\Delta_{\rm decay}+\Delta_{\rm LTTE}$) are indicated by the green dotted, blue dashed, and red solid line, respectively.}
  \label{fig2_data_model}
\end{figure}

An illustrative example of the measurement for the individual mid-eclipse time is shown in Figure \ref{fig_etv}. The big red dots represent the observed data of an arbitrarily eclipse. Here, we select the orbital epoch 53. The flux at the mid-eclipse time (blue diamond) is the minimum of the polynomial fitting (green line) for those big red dots. The uncertainty of $t_0$ is directly obtained from the fitting procedure. The observed data that are available for fitting the secondary eclipse is about 3 times more than those for the primary eclipse. A crude estimate shows that the uncertainty of the $t_0$ for the secondary is only $\sim$2.3 times larger than that of the primary eclipse, even though the primary eclipse is $\sim$8 times deeper. The time system of the mid-eclipse is clearly documented, as shown in Figure \ref{fig2_data_model}.

\begin{figure}
\centering
  \includegraphics[width=0.5\textwidth]{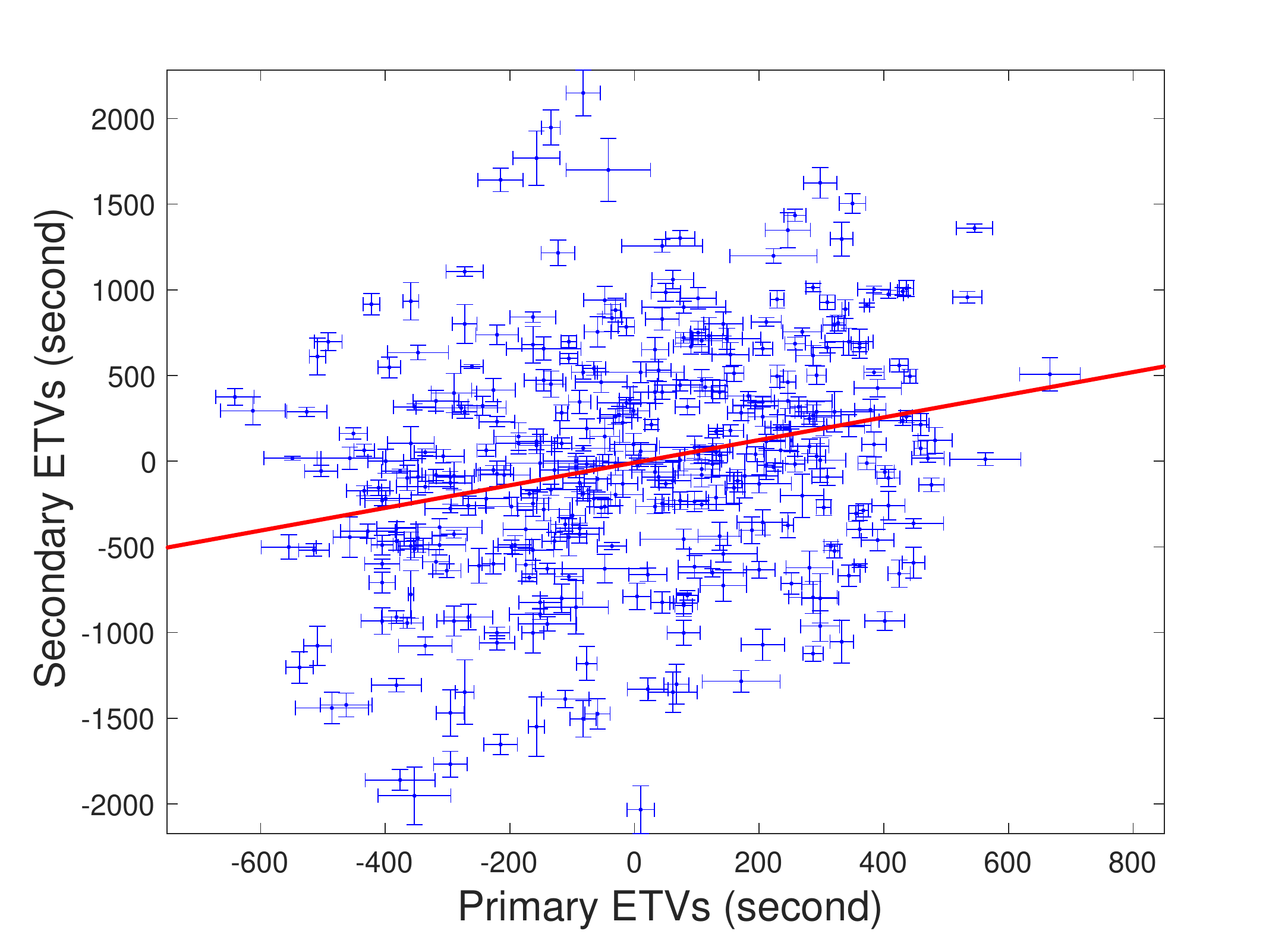}
  \caption{The correlation of ETVs between the primary and secondary. The positive correlation excludes the phenomenon of apsidal motion and star spots in KIC 3766353. Note that the scale of the abscissa and the ordinate is different.}
  \label{fig3}
\end{figure}

\subsection{The correlation of timing variations}\label{sec4.2}

Previous studies noticed that the timing variations formed from the mid-times of the primary and secondary eclipse show correlated or anti-correlated relations. The anti-correlated behaviour can be explained by the apsidal motion effect \citep{Ste1939, Cla1993}, or the presence of star spots on the surface of the host start \citep{Tra2013}. On the other hand, the positive correlation may be attributed to a hierarchical multiple system \citep{Con2014} or orbital period decay \citep{Pat2017, Wil2017,Yee2020}. 

We adopt a correlation coefficient to estimate the relation between the primary and secondary timing variations. As seen in Figure \ref{fig3}, the results show a positively correlation between the mid-times, which implies that there is likely  a tertiary companion and/or orbital decay in KIC 3766353. 

The apsidal motion effect is often seen in eccentric binary systems. However, we don't see clear evidence of anti-correlation between the mid-times, and hence the apsidal motion effect and the presence of star spots are excluded from the following analysis. Practically speaking, distinguishing between different physical processes is a key to search for orbital decay, and we will discuss the details in the next section.

\section{Variations of orbital period}\label{sec5}

The observed timing variation may be a combination of the presence of a hidden tertiary object and orbital decay. Neither of these phenomena is mutually exclusive. Fortunately, timing variations affected by different physical processes act on various time-scales with various amplitudes, according to which we may differentiate bewteen the contributions of each process. In this section, the analytic approach of the orbital decay, the LTTE effect, and the dynamical perturbation are discussed in Section \ref{sec5.1}, Section \ref{sec5.2}, and Section \ref{sec5.3}, respectively. The reason why the dynamical perturbation can be ignored is also discussed in Section \ref{sec5.3}. Finally, we draw the implications of ETVs via Markov chain Monte Carlo (MCMC) simulations in Section \ref{sec_add6}.

\subsection{Orbital decay}\label{sec5.1}

The model of orbital decay assumes the orbital period to be changing uniformly with time \citep{Pat2017, Yee2020}:

\begin{equation}\label{equ_dec}
\Delta_{\rm decay} = c_0 + c_1E + \frac{1}{2}\frac{dP}{dE} E^2 \ ,
\end{equation}
where $E$ is the orbital epoch, and the coefficients $c_0$ and $c_1$ give corrections to the reference epoch and the average eclipsing period, respectively. ${dP}/{dE}$ is the rate of orbital decay in units of day per cycle (d/c).

In theory, the orbital decay can be modelled by a quadratic polynomial. For instance, the well-known exoplanet WASP-12b is the first hot jupiter that is confirmed with orbital decay using the quadratic polynomial fitting method \citep{Yee2020, Tur2021}. However, there are several physical processes, such as the mass exchange between the components, wind-driven mass loss, tidal dissipation and magnetic braking, can result in a quadratic variation in the orbital period decay.

\subsection{Light-travel time effect}\label{sec5.2}

To study the orbital variation of KIC 3766353, we adopted the Light-travel time effect (LTTE) hypothesis \citep{irw1959}. LTTE assumes that the variation of the eclipse time of the binary components is due to the presence of a hidden distant companion, hence in a hierarchical multi-star system. As the binary pair orbit around their barycentre, the third companion also moves around the common barycenter with the eclipsing pair in an outer orbit. As a result, the eclipse of the binary pair occurs sometimes delayed and sometimes earlier, depending on the actual position of the stars and the observer. The form of LTTE can be written as a function of the outer orbital configuration \citep{Bor2016}:

\begin{equation}\label{equ_ltte}
\Delta_{\rm LTTE} = -\frac{a_{\rm AB} \sin i_2}{c} \left[  \sqrt{1-e_2^2} \sin E_2 \cos \omega_2 \\
  + (\cos E_2-e_2) \sin \omega_2 \right] \ ,
\end{equation}
where $a_{\rm AB}$ is the semi-major axis of the absolute orbit that is related to the ratio between the mass of the hidden body $m_{\rm C}$ and that of the triple system $m_{\rm ABC}$ as $a_{\rm AB} = (m_{\rm C}/m_{\rm ABC}) a_2$. The parameters of the outer orbit are denoted by the subscript ``2", including the relative semi-major axis ($a_2$), orbital inclination ($i_2$), eccentricity ($e_2$), eccentric anomaly ($E_2$), argument of periastron ($\omega_2$), and $c$ is the speed of light. The amplitude of the LTTE contribution is given approximately by: 

\begin{equation}
\mathcal{A}_{\rm LTTE} = 1.1 \times 10^{-4} f(m_{\rm C})^{1/3} P_2^{2/3} \sqrt{1-e_2^2 \cos^2 \omega_2} \ , \label{equ_altte}
\end{equation}
where the mass function $f(m_{\rm C})$ is defined as

\begin{equation}
f(m_{\rm C}) = \frac{m^3_{\rm C} \sin^3 i_2}{m^2_{\rm ABC}} = \frac{4 \pi^2 a^3_{\rm AB} \sin^3 i_2}{G P^2_2} \ ,
\label{equ_fmc}
\end{equation}
$G$ is the gravity constant. In regard to units, the deviation amplitude and period are in days and the masses are in solar mass $M_\odot$. Similar to radial velocity observations, the mass function $f(m_{\rm C})$ does not allow either the inclination of the outer orbit $i_2$ or the mass of the tertiary component $m_{\rm C}$ to be uniquely deduced.

\subsection{Dynamical perturbations}\label{sec5.3}

In a binary systems with a third companion, the orbital motion no longer remains purely Keplerian because all six orbital elements are affected by perturbations. The dynamical perturbations may dominate over contributions caused by LTTE, if the third companion has a strong interaction with the inner binary. Therefore, we estimate the amplitude of the dynamical perturbations to determine whether the contribution of dynamical perturbation needs to be taken into account.

For eccentric inner binaries with period $P_1$, the amplitude of the dynamical contribution takes the expression \citep{Bor2015, Bor2016}

\begin{equation}
\mathcal{A}_{\rm dyn} = \frac{15}{16 \pi} \frac{m_{\rm C}}{m_{\rm ABC}} \frac{P^2_1}{P_2}(1-e^2_2)^{-3/2} \ . \label{equ_adyn}
\end{equation}

The ratio of amplitudes $\mathcal{A}_{\rm dyn}/\mathcal{A}_{\rm LTTE}$ tells which effect is dominant. A ratio less than $\sim$0.1 indicates that a pure LTTE solution is sufficient to reproduce the period variations and the effect of dynamical perturbations is negligible.

\begin{figure}
\centering
  \includegraphics[width=0.5\textwidth]{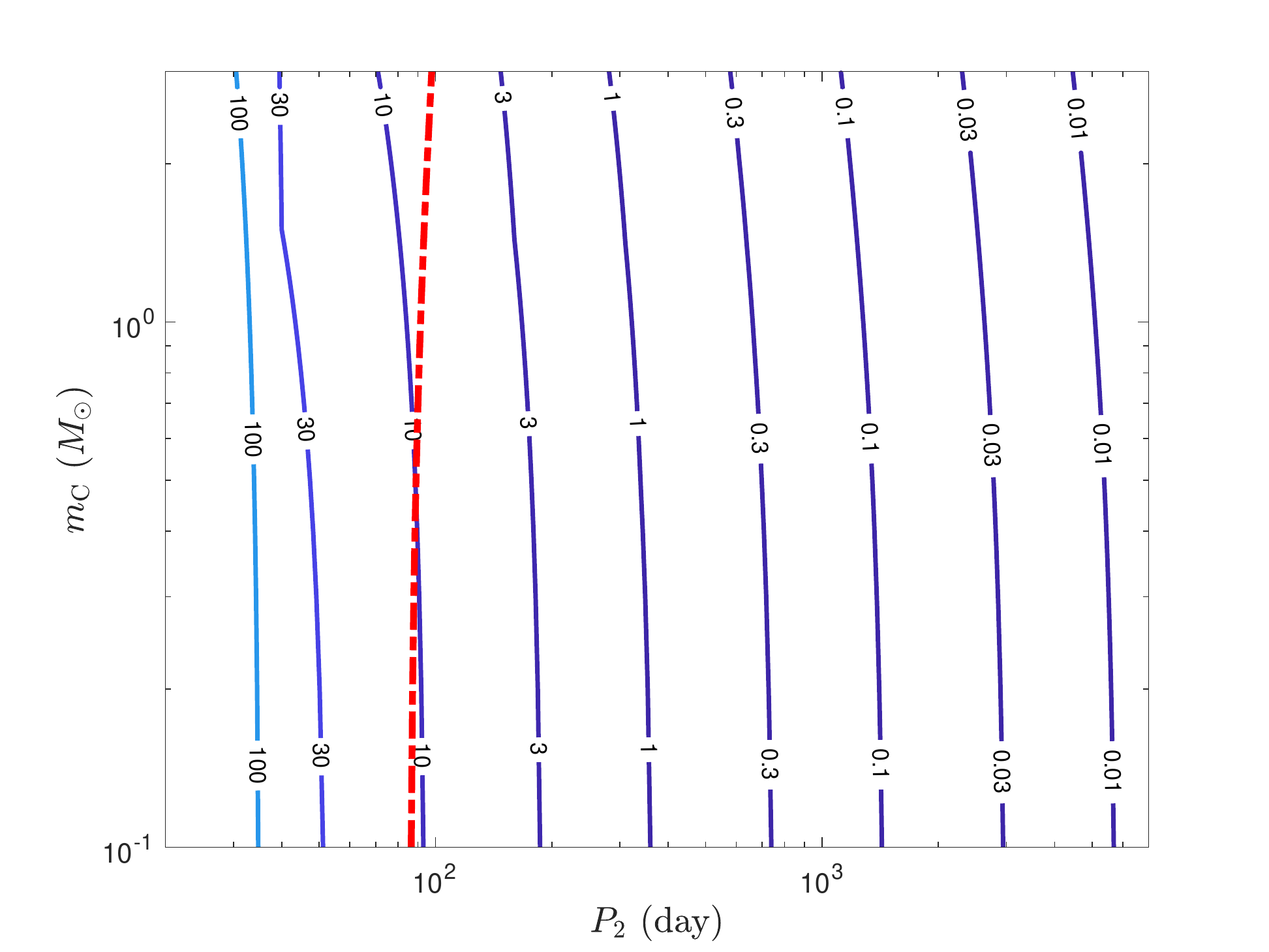}
  \caption{The ratio of the amplitude of dynamical perturbations $\mathcal{A}_{\rm dyn}$ given by Equation (\ref{equ_adyn}) and the LTTE amplitude $\mathcal{A}_{\rm LTTE}$ given by Equation (\ref{equ_altte}). A series of given ratios (values given by the numbers) are illustrated by the lines. The long-term dynamically stable systems expected to be lie to the right-hand side of the red dot-dashed line, see Equation (\ref{equ_p2}). The dynamical perturbations can be securely ignored, if the ratio $\mathcal{A}_{\rm dyn}/\mathcal{A}_{\rm LTTE}$ is less than $\sim$0.1.}
  \label{fig_dyn}
\end{figure}

~\\

Indeed, the mutual orbits of a triple system are expected to have long-term dynamical stability that can be conveniently tested by a criteria expressed in terms of the outer orbital period \citep{Rap2013} as

\begin{equation}
P_2 \gtrsim 4.7 \left ( \frac{m_{\rm ABC}}{m_{\rm AB}} \right )^{1/10} \frac{(1+e_2)^{3/5}}{(1-e_2)^{9/5}}P_1 \ . \label{equ_p2}
\end{equation}

We assume $e_2 = 0.5$, $i_2 = 60^\circ$ for the outer orbit, and $m_{\rm AB}=1.11 \ M_{\odot}$ for the sum of the masses of the inner two binary stars. Figure \ref{fig_dyn} illustrates the amplitude ratios of $\mathcal{A}_{\rm dyn}/\mathcal{A}_{\rm LTTE}$ as a function of the outer orbit period $P_2$ and  the mass of the third companion $m_{\rm C}$. Dynamically stable systems would be lie to the right-hand side of the red dot-dashed line. As could be expected from the analytical forms of Equations (\ref{equ_altte}) and (\ref{equ_adyn}), the LTTE dominates for longer orbital periods and massive tertiary objects of the triple system, while the dynamical perturbation dominates for the third body that tightly closed to the inner binary. In practice, the dynamical perturbations can be securely ignored when the ratio $\mathcal{A}_{\rm dyn}/\mathcal{A}_{\rm LTTE}$ is less than $\sim$0.1, because the contribution of dynamical perturbations is fairly small.

\subsection{Model solution}\label{sec_add6}
We used the analytic approach as described above, i.e. Equaction (\ref{equ_dec}) and/or Equaction (\ref{equ_ltte}), to fit the ETV curves, shown in Figure \ref{fig2_data_model}, for both of the primary and secondary eclipses simultaneously. Three kinds of models are considered: pure model of orbital decay, pure model of LTTE, and a combined model of orbital decay and LTTE. 

Considering that the LTTE effect fitted is a sinusoid with period equal to or longer than the data span, the quadratic period decay term may simply be fitted by a subharmonic of this signal. We use a nested sampling \citep{spe2020} procedure to assess the Bayesian evidence for the orbital decay hypothesis. Multi-Core Markov-Chain Monte Carlo \citep[mc3,][]{cub2017}\footnote{https://mc3.readthedocs.io/en/latest/}  supported for nested sampling is applied to estimate the best-fitting values and the credible regions for the parameters for the three models.

The fitting parameters might converge incorrectly to a local minimum, if an inadequate exploration is performed for the parameter space. Therefore, we expand the range of values and set all parameters as free in our adjustment process. For the three models, we constrain the best-fitting parameters by minimizing $\chi^2$: 

\begin{equation}
\chi^2 = \sum_{i=1}^{N_{\rm dp}}\left( \frac{f_{{\rm obs},i} - f_{{\rm mod},i} }{\sigma_i} \right)^2 \ ,
\end{equation}
where $N_{\rm dp}=763$ is the number of data points, $f_{{\rm obs},i}$ is the observed data with uncertainty $\sigma_i$, and $f_{{\rm mod},i}$ is the simulation data of the model.

The best-fitting models of the pure orbital decay, the pure LTTE and the combined model are marked with green dotted, blue dashed, and red solid lines, respectively, as shown in Figure \ref{fig2_data_model}. The values of $\chi^2$ are listed in Table \ref{tab2}. We find that the combined model fits the observed ETV data best.

In order to see if any additional free parameter is necessary to explain the data. A robust way to compare models with different number of parameters is the Bayesian Information Criterion: 

\begin{equation}
{\rm BIC} = \chi^2 + k\ {\rm log}\ N_{\rm dp} \ ,
\end{equation}
where $k$ is the number of free parameters. The preferred model is the one that produces the lowest BIC value. The statistical significance of the BIC is the penalty for a higher number of fitted parameters. Table \ref{tab2} also gives the BIC values of the three models. Again, the BIC favors the combined model over pure orbital decay model by $\Delta \rm{BIC}_1 = 10319.8$, and over pure LTTE model by $\Delta \rm{BIC}_2 = 172.6$. Therefore, the combined model is overwhelming in the preferred interpretation of the observed timing variations of KIC 3766353.

The physical implications of the hierarchical triple system are shown in the last column of Table \ref{tab2}. The hidden third body is a low-mass component with the lower-limit mass of $\sim 0.26 \ M_{\odot}$ at a distance of $\sim 111.4\ R_{\odot}$ from the inner binary. The lower-limit mass is estimated from the mass function $f(m_{\rm C})$, equation (\ref{equ_fmc}), after adopting the inclination of the outer orbit $i_2$ as $90 ^{\circ}$. The outer orbit has period $P_2 \sim 1416.85$ d with eccentricity $e_2 \sim 0.48$ and argument of periastron $\omega_2 \sim 129.2^{\circ}$. In this case, the amplitude and shape of LTTE can reproduce the observed timing variations. The best fitting result also indicates that the inner orbit of the triple system has an orbital period decay rate as $\dot{P_1} \sim 358$ ms yr$^{-1}$.

\begin{table*} 
\renewcommand\arraystretch{1.5}
  \caption{ Best-fitting parameters for the three analytic models.}\label{tab2}
  \begin{center}
  \begin{tabular}{cccc}
  \hline
Parameters & Orbital decay model & LTTE model & Combined model \\
\hline
$c_0$ (second) & $-222.07_{-2.62}^{+2.58}$  &  / & $-190.94_{-56.85}^{61.85}$  \\
$c_1$ (second) & $1.59 \pm -0.01$  & /  & $0.67_{-0.09}^{+0.07}$  \\
$\dot{P_1} \ {\rm (ms\ yr^{-1})}$ & $903.73_{-3.71}^{+3.68}$ & / & $358.04_{-53.72}^{+45.74}$\\
$a_{\rm AB}\sin i_2\ (R_{\odot})$ & / & $151.56\pm 0.50$ & $111.40_{-10.17}^{+17.98}$\\
$e_2$ & / & $0.22\pm 0.004$ & $0.48_{-0.30}^{+0.29}$\\
$P_2\ {\rm (day)}$ & / & $1556.23_{-4.72}^{+4.92}$ & $1416.85_{-38.16}^{+41.83}$\\
$\omega_2 \ (^\circ)$ & / & $130.78_{-0.44}^{+0.47}$ & $129.18_{-5.81}^{+8.67}$\\
\hline
 \multicolumn{1}{c}{ } & \multicolumn{3}{c}{criteria for model selection} \\
 \hline
$\chi^2$ & 91959.6 & 81812.4 & 81639.8 \\
BIC & 91968.2 & 81823.9 & 81660.0 \\
\hline
\end{tabular}
\end{center}
\end{table*}

\section{Summary and Conclusions}\label{sec6}

In the theoretical framework, orbital shrinkage raised in the close binary system requires the angular momentum transfer from the orbit to the stars, causing  the tidally excited oscillations in the primary companion as one of the results. The straightforward idea is that the prototype of tidally excited oscillations, i.e. heartbeat stars, appear to merit a high priority for searching for potentially detectable orbital shrinkages. KIC 3766353 is one of the heartbeat stars with tidally excited oscillations, and the primary and secondary eclipse time variations imply the change of orbital period in this system. 

By using the latest version 2.3 of PHOEBE, we obtain the orbital and stellar parameters given by the \textit{Kepler} observed light curve. The mass and radius are $0.76^{+0.21}_{-0.13}\ M_{\odot}$, $ 1.32^{+0.16}_{-0.06}\ R_{\odot}$ for the primary star, and $0.35_{-0.16}^{+0.13}\ M_{\odot}$, $0.34_{-0.03}^{+0.04}\ R_{\odot}$ for the secondary star, respectively. That is to say, the system is composed of a subgiant primary and a red dwarf companion. The tidal friction of subgiant stars becomes orders of magnitude larger than the main sequence (MS) stars. The subgiant with surface convective zones effectively converts orbital angular momentum via tidal interactions induced by the massive companion. The configuration of the system and the presence of tidally excited oscillations  suggest that the orbital decay is apparent in KIC 3766353.

We have measured the primary and secondary eclipse time variations that have allowed us to investigate this system more correctly, because some degeneracies in the dynamical parameters might break, such as apsidal motion \citep{Rap2013} and octupole terms \citep{Bor2015}. Finally, the ETV curves for both of the primary and secondary are simultaneously fitted in three analytic models: pure orbital decay model, pure LTTE model, and a combined model of orbital decay and LTTE. Based on the sufficient parameter space of MCMC simulations with nested sampler, the criteria ($\chi^2$ and BIC) for model selection are overwhelming that the combined model perform better than the pure orbital decay model and the pure LTTE model to interpret the observed timing variations of KIC 3766353. The combined model implies that there is a third companion with a lower-limit mass of $\sim 0.26 \ M_{\odot}$ at a distance of $\sim 111.4\ R_{\odot}$ from the inner binary. The inner binary system consists of a subgiant and a red dwarf that has an orbital decay rate as $\sim 358$ ms yr$^{-1}$. Future observations are expected to provide precise timing monitors to delve deeper into this topic.

\end{CJK}

 \section*{Acknowledgements}

We acknowledge the anonymous referee for his/her suggestions that greatly improved this paper. O.J.W. is grateful to Dr. Xiaoling Yu for the technical help, and Yong-Hao Wang, and Shang-Fei Liu for helpful discussions. This work is supported by the National Key R\&D Program of China (No. 2020YFC2201200), the National Natural Science Foundation of China (No. 11803012), the science research grants from the China Manned Space Project (No. CMS-CSST-2021-B09), the Fundamental Research Funds for the Central Universities: Sun Yat-sen University (No. 20lgpy174) and Nanjing University, and the Youth Science and Technology Talents Development Project of Guizhou Education Department (No. KY2018421). Y.C. has been supported by the National Natural Science Foundation of China (No.  11373064, 11521303, 11733010, and 11873103), Yunnan Natural Science Foundation (No. 2014HB048), and Yunnan Province (No. 2017HC018). J.C. is supported by a grant from the Max Planck Society to prepare for the scientific exploitation of the PLATO mission.

\section*{DATA AVAILABILITY}
Data and source code are available upon reasonable request to the corresponding author.



\bsp	
\label{lastpage}
\end{document}